\documentclass[runningheads]{svmult}

\usepackage{makeidx}   
\usepackage{graphicx}  
\usepackage{subeqnar}  
\usepackage{multicol}  
\usepackage{physprbb}  
\makeindex             

\begin{document}
\title*{The Abundance of Elements in Cool Stars, as Determined from High-Resolution, 1-5\,$\mu$m Spectroscopy}
\toctitle{The abundance of elements in Cool Stars
}
%
%
\titlerunning{The abundances of elements in Cool Stars}
%
\author{N. Ryde \and B. Gustafsson \and K. Eriksson \and R. Wahlin}
\authorrunning{Nils Ryde et al.}
%
%
\institute{Department of Astronomy and Space Physics, Uppsala
University, Sweden}

\maketitle              

\begin{abstract}

We review the field of abundance determinations of elements in cool stars,
with special interest paid to determinations based on analyses
of high-resolution, $1~-~5\,\mu$m spectra. We discuss the current
status, problems, and challenges of exploring high-resolution,
near-infrared spectra. In particular, advantages and drawbacks
are pointed out. A few examples of current,
chemical-abundance determinations are high-lighted and, finally, we discuss
the development and future prospects of the field.

\end{abstract}

\section{Introduction}

When P. Connes in the years 1966-67 constructed and set up his prototype Fourier Transform
Spectrometer (FTS) for infrared studies at Observatoire
de Hautes Provence, his main objective was not to study stars. In his
review on "Astronomical Fourier Spectroscopy" \cite{connes}
he writes "The spectra of a few bright stars were also
recorded on a spare-time basis." Most of the observing time was instead
devoted to
planetary studies. The stellar spectra obtained were described by Spinrad et
al. \cite{spinrad}.
Among the results described is a first
determination of the isotopic ratio $^{12}$C/$^{13}$C for Betelgeuse. Later Maillard \cite{maillard}
obtained the isotopic ratios $^{12}$C/$^{13}$C, $^{16}$O/$^{17}$O and $^{16}$O/$^{18}$O for Alpha Her from infrared
CO lines with a new FTS at Observatoire de Hautes Provence, and in collaboration
with Flaud and Camy-Peret \cite{flaud}, he made the first
estimate of the H$_2$O abundance in R Cas.
In a report on the status of infrared FTS spectroscopy, Maillard
\cite{maillard:2}
presented a table of all astronomical FTS:s in Europe and USA at that time, and the
astronomical objects observed. It is interesting, but not unexpected,
that the favorite
stellar sources were bright infrared sources like M stars, carbon stars, miras,
with very complex molecular spectra. This work also stimulated work in
molecular spectroscopy in order to interpret the wealth of data further.
Also with the Fourier-transform spectrometers used at McDonald Observatory and Kitt Peak
National Observatory (KPNO), early studies were devoted to isotopic ratios, notably
of carbon and oxygen isotopes \cite{barnes,dom,hls}.
Another area of intense study of stars was the exploration of the dynamics of mira atmospheres by
Hinkle and collaborators. An early attempt to derive
abundances of elements (C, N, and O) by high-resolution, infrared spectroscopy was made for
Betelgeuse by Lambert et al. \cite{lambert:84} in 1984, soon followed by the determination of
C, N, and O abundances and $^{12}$C/$^{13}$C ratios for 30 Galactic carbon stars by Lambert et
al. \cite{lambert:86} in 1986. By then, high-resolution spectroscopy in the infrared  and the
analysis of the spectra produced had matured to become
central tools for studying cool-star atmospheres, and not least for
measuring chemical abundances. (For a more detailed account, see, for example, Gustafsson \cite{bg:89}.) A great step
forward was next taken with the
development of cryogenic echelle spectrometers such as the Phoenix spectrometer
at KPNO and later at Gemini Observatory, which made it possible to reach much fainter stars
\cite{pila}.

The art of determining chemical abundances in cool
stars thus benefits strong\-ly
from the realization of sensitive spectrographs capable of providing
high-resolution, near-infrared spectra, such as the Phoenix and the CRIRES
spectro\-graphs. In this review we will examine why abundances of elements in
cool stars are of interest, why we need high-resolution
spectroscopy, and discuss the advantages and drawbacks of
high-resolution, 1-5 $\mu$m spectroscopy.  The status of model
atmospheres needed in the analyses based on spectral synthesis
will also be reviewed briefly.
By cool stars we mean K and M dwarfs, and M, C, and S giants,
i.e. red giants including AGB stars. By high-resolution spectra, we
will denote spectra dispersed by a resolving power, $R$, greater than $50\,000$.

\section{Why do we need chemical abundances in cool stars?}

Accurate abundances\footnote{For some elements and purposes,
for example in studies of Galactic chemical evolution, relative abundances need to be determined
at an accuracy of 0.1 dex or better} of atoms, their isotopes, and molecules
in cool-star atmospheres are necessary for the understanding of the stars
themselves, the interpretation of their spectra in terms of fundamental
stellar parameters, the exploration of their interior nuclear reactions and
mixing processes, and of the structure of their outer layers and the driving
mechanisms of their winds. Also,
accurate abundance patterns in cool stars are required for the
understanding of late stellar evolution. In order to put constraints on
evolutionary models, the details of, for example, the processes taking place during the
transition phases from the Asymptotic Giant Branch (AGB) to the Planetary
Nebula
phases as reflected in changing surface abundances, can and should be
investigated observationally. Also, stellar evolution for different initial
conditions can be probed by measuring chemical abundances.

Furthermore, theoretically calculated yields of elements in different
evolutionary phases of stars are needed in Galactic evolution models, but are,
in several cases, still very uncertain. Direct measurements of
yields in these phases, such as the AGB phase and the Planetary Nebulae
phase, are therefore essential (cf. \cite{bg:arnould}).
The AGB stars are thought to be the main
contributors of the s-elements (e.g. Sr, Y, Tc, Ce, and Hg) and significant
contributors of $^{19}$F and $^{12}$C \cite{bg:ryde}. A
systematic observational study of how these yields vary with initial stellar
composition and mass is an important programme, far from completion.

Moreover, K and M dwarfs and sub-giants are also useful as probes
of the chemo-dynamic evolution of the Milky Way
system, since their surface abundances reflect the composition of the
gas clouds in which they were once formed. Also K and M giants are used as
valuable probes, e.g. in regions which may not be readily accessed with
dwarfs or sub-giants such as the Galactic Bulge or external galaxies such as
the dwarf spheroidals. If chemical abundances in giants could be measured as
accurately as for solar-type stars, and
be understood theoretically, e.g. in terms of the modifications of the
initial stellar abundances by the individual stars, their value as probes
would be even greater. To a considerable degree, such further understanding
relies on more systematic abundance measurements. Clearly, abundance studies
of elements in cool stars are of high scientific value.



\section{The virtue of high-resolution spectroscopy}

Since cool stars have crowded spectra due to the presence of
molecules, high-resolution spectrographs, such as CRIRES, are
needed in order to disentangle the spectral information. The cooler the star
the more crowded the spectrum. For example, for a typical cool,
oxygen-rich red-giant of $T_{\mathrm{eff}}= 3\,000\,\mathrm{K}$, and
$\log g = 0.0$, the general flux distribution follows the
H$^{-}_\mathrm{bf}$ and H$^{-}_\mathrm{ff}$ continuous opacities
leading to a maximum continuum level at approximately $1.6\,\mu$m.
However, molecular lines determine the detailed flux distribution
of the star. Whereas the atomic absorption does not play a large role anywhere
in the spectrum except in the extreme blue end, TiO and VO totally
dominate in the optical region and water vapour, OH, CO, and SiO dominate
the near-infrared region. For a corresponding cool, carbon-rich
red giant, the infrared spectrum is dominated by CN, CO (which
dominate the M-band), CH, C$_2$, and the polyatomic molecules
C$_3$, C$_2$H$_2$, and HCN, the latter two apparent in the Johnsson
L band. It is clear that in order to model spectra which are so
dominated by molecules, even for high-resolution spectroscopy,
spectral synthesis is a prime tool for analysis.

A further merit of high-resolution spectroscopy is the match
between the resolution and the intrinsic widths of stellar
spectral lines. Lines formed in a stellar atmosphere have an
inherent line width of the order of $2 - 5\,\mathrm{km\,s^{-1}}$.
In these crowded spectra, a resolution of approximately $R \sim 100\,000$ is often needed to
obtain full information from the spectra about line strengths. High spectral-resolution also makes it possible to detect
and decompose line blending, large-scale
motions, and acquire more precise knowledge on, for example, `turbulence' and
magnetic-field strengths.

\section{Advantages of an abundance analysis at $1-5\,\mu$m}

There are a number of advantages of studying chemical abundances in cool
stars at near-infrared wavelengths. Here, we will mention a few.

A general reason for studying cool stars at
$1-5\,\mu$m is their brightness at these wavelengths. A
red giant of $T_{\mathrm{eff}}= 3\,000\,\mathrm{K}$, and $\log g =
0.0$ has its largest flux output around $1-2\,\mu$m, i.e. in the
Johnson J and H bands.  Red giants
are also bolometrically luminous. (Actually, a majority of all stars that can
be seen, even optically, on a dark night are red giants.)

Furthermore, at infrared wavelengths the opacity of interstellar or circumstellar
dust is less than in the optical domain. Thus, owing to the fact that infrared
radiation can penetrate through optical dust-obscuration, stars
also in dusty environs of the Universe, for instance star-forming
regions, or stars in the Galactic Bulge with a large column density of dust
in the line-of-sight, can readily be observed.

The fact that the opacity in a cool, stellar atmosphere
has its minimum at $1.6\,\mu$m implies that the depth of
formation of the continuum is largest at $1.6\,\mu$m. Weak,
infrared atomic lines in the H band are therefore mostly formed deep in the atmosphere
where the physical state is relatively well known. This is a definite
advantage. Moreover, in the Rayleigh-Jeans regime, the intensity is
less sensitive to temperature variations; just as for black-body radiation
($B_\nu(T)$), $\delta B_\nu(T) / \delta T$ is small in the Rayleigh-Jeans
regime. This means that the
effects of, for example, effective-temperature uncertainties or surface inhomogeneities on
line strengths may be smaller in the infrared. As an example, we can compare iron abundances
derived from a given, weak spectral-line
from neutral iron in two cases: In the first case the iron abundance is derived from a combination of
50\% of a giant-star model-atmosphere with an effective temperature of $T_{\rm eff} = 3\,000\,$~K and
50\% of a $T_{\rm eff} = 4\,000\,$~K atmosphere, thus simulating an inhomogeneous
atmosphere with two temperature components. In the second case the abundance is derived from a
homogeneous model with the same total flux as the two-component model, i.e. an atmosphere with a $T_{\rm eff} = 3\,600\,$ K.
If the weak Fe{\sc i} line is situated at $2\,\mu$m  then the derived abundance would only be approximately 0.05
dex higher from the two-component model than from the homogeneous model for
high-excitation lines. For the low-excitation lines we get almost no effect. If,
on the other hand, the weak line were situated in the optical wavelength region, say at 500 nm,
the effect would be larger, up to 0.2~dex, and now for the low-excitation lines. For the
high-excitation lines the effect is smaller and even reversed.

A further advantage is the fact that in the $1-5\,\mu$m domain, lines from most molecules are often pure
vibration-rotational lines. However,
this is not the case for CN and C$_2$. The forest of molecular lines is also cleaner
in the sense that several electronic systems less often overlap severely compared to the ultraviolet.
Since the transitions
occur within the electronic ground-state, the assumption of Local
Thermodynamic Equilibrium (LTE) in the analysis of the molecules is
probably valid \cite{hinkle_lambert:75}. For the warmer of the cool
stars, an additional advantage of analyzing the near-infrared
compared with the optical wavelength region, is that it is easier
to find portions of the spectrum which can be used to define a continuum. In
fact, also the number of atomic and ionic lines is much smaller than in the
ultraviolet, reflecting the sparser term diagrams for the lower, most-populated states of the atoms. This is,
however, also a general drawback of the infrared spectral region - the number of useful atomic and ionic lines
is relatively limited in practice.

Lines from molecular species (with a host of lines in the $1-5\,\mu$m region) are beneficial in the
sense that their identification is relatively easy due
to an (often) regular multiplicity of lines. For complex molecules, however, the easy identification is
true only if there exists a theoretical line list of that
specific molecule which is sufficiently accurate (see for example
the discussion on the identification of water vapour in
Arcturus \cite{journ2}). The ladder of different vibration-rotational transitions,
with varying line strengths, is also
suitable for a semi-empirical modelling of an atmosphere, since the different lines
together sample a large range of atmospheric, optical depths.

Measurements of isotopic abundances of carbon, nitrogen, and
oxygen are of great interest, for example, for the study of
nucleosynthesis and stellar evolution. The infrared wavelength
region is ideal for studying isotopic abundances, chiefly since
the isotopic shifts are larger for molecules than for atoms, and
molecular lines are ubiquitous in the infrared. (The mass
difference of two isotopic molecules leads to unequal rotational
constants, affecting the wavelengths of rotational and
vibration-rotational lines of the isotopic molecules.) For example,
the isotopic shift between the $^{12}$CO($v=1-3$) and
$^{13}$CO($v=1-3$) band heads at $2.4\,\mu$m  is as much as
$0.05\,\mu$m.

A further benefit of working with molecules is that for an element
which can form many different molecules, such as carbon, nitrogen
and oxygen, several diagnostics can be utilized for the
determination of the abundance of that element.

Finally, the formation and destruction of molecules can be very
sensitive to the temperature structure in the atmosphere. Thus, for
certain cases the molecular lines may probe the upper-most
atmosphere just because they exist only there. This is of diagnostic
value for the study of atmospheric surface layers, but is a drawback
in abundance analyses which one would wish to be as insensitive as possible to
the uncertainties in the atmospheric structures.

\section{Drawbacks of an abundance analysis at $1-5\,\mu$m}

There are obviously also some disadvantages in using the
infrared to study abundances in cool stars, but none of these is
generally severe. Many cool stars form
dust in their outer atmospheres, in a few cases so much that the
dust obscures the star. The main spectral contribution of the dust
is, however, an additional spectrally broad, thermal emission on top of the stellar spectrum.  The maximum of this emission
lies at $3$ to $12\,\mu$m, depending on the temperature of the dust, which is
typically a few hundred to a thousand Kelvin. This extra emission
complicates the spectral analysis. Also, the telluric atmosphere, with
its ubiquitous water-vapour lines and its thermal emission beyond 2 microns hampers the analysis, as does
the thermal emission from the telescope.

Moreover, as was mentioned above, there are much fewer atomic and
ionic lines in the infrared region. The ones present often originate
from highly excited levels in atoms, which also complicates
an interpretation. Furthermore, many lines are not properly
identified and/or lack known oscillator strengths, which are
needed in an abundance analysis. A further drawback for a molecular
abundance analysis in this wavelength region is the lack of
clear and well studied signatures from several molecules, such as TiO and ZrO beyond the J band.

Finally, even though significant advances have been made in
the technology for recording near-infrared light, existing
spectrometers are still much less effective than optical ones, one
of the main reasons being the present lack of cross-dispersion.

Thus, to summarize, the advantages of chemical-abundance analyses of cool-star atmospheres in the infrared
are considerable and numerous.  The disadvantages are few, and some of these will
even be possible to surmount in the near future.

\section{The status of high-resolution abundance analyses at $1-5\,\mu$m}

In order to analyze stellar spectra, a model atmosphere must
be calculated. Subsequently, a synthetic spectrum is computed by
solving the radiative transfer through this model atmosphere. How
realistic are cool, stellar model-atmospheres and synthetic spectra?
Here, only a short account will be given. For more detailed discussions, see,
for example \cite{bg_jorg} or \cite{bg_hoefner}.

 \subsection{The status of stellar atmospheres}

The answer to the question on how well model atmospheres reflect reality, has
two components. First, the input parameters have to be well
determined. A star's fundamental parameters are its effective
temperature, $T_\mathrm{eff}$, surface gravity, $\log (g)$,
metallicity, and mass (or radius). Second, the physical
description of the star's light-forming atmosphere has to be
relevant, i.e. all important physics has to be included at a
sufficiently high level of realism.

Concerning the stellar parameters, in general, the
determination of a star's effective temperature, $T_\mathrm{eff}$,
is satisfactory also for cool stars based on their angular diameters (for nearby giants
and supergiants) and on the InfraRed Flux Method (IRFM).
The assessment of a star's $\log (g)$ is very
difficult, with uncertainties of the order of $\pm
0.3\,\mathrm{dex}$ or even more for cool giants. However, the surface gravities of stars in the
Magellanic Clouds and in the Dwarf Spheroidals in the Local Group
can be assigned more readily because of their relatively well-known distances. The
astrometric GAIA mission will contribute to the solution of this problem by providing parallaxes to
a host of stars in the Galaxy. The determination of the overall-metallicities
of cool stars may be difficult due to the forest of molecular lines
and few pure atomic lines. The determination of masses is even more involved.
However, since measurements of diameters of nearby giants are
possible by means of interferometry, a mass can in principle be calculated from
this diameter and the spectroscopic surface gravity of the star. Thus, although several
severe difficulties do exist in the determination
of a cool star's fundamental parameters, these difficulties are not insurmountable in principle.

Concerning the model atmospheres, several ingredients are
important. The physical structure of a model depends on accurate
opacities. For cool stars these are dominated by lines from
diatomic and small, polyatomic molecules. Today, the observed line-blocking in M
giants and dwarfs is well reproduced by models (see, for example, \cite{Alvarez}).
However, the water
opacity is still in need of improvement (see, for instance,
\cite{jorg},
\cite{jones}, and
\cite{rdor}).
The modelling of carbon stars is still hampered by the lack of reliable opacity
data of some polyatomic molecules, such as C$_2$H$_2$,
CH$_4$, C$_2$H, and C$_3$H.
Important work on carbon-star
opacities is being performed by several groups, see, for example,
\cite{harris} and
\cite{borysow}.

For bright giants and supergiants, which have atmospheres with large geometric extensions,
the sphericity affects
the radiative transfer and therefore the physical structure. For
these stars, codes for calculating models taking the spherical nature of a star into account, do exist and
should be used in the analysis (see, for example,
\cite{plez} and
\cite{hau} and others). Non-static (i.e. dynamic)
models for red giants, Asymptotic Giant Branch stars, and red
supergiants are still under development, but have already yielded
several important qualitative results. Dynamic models are
crucial for the understanding of these types of stars, as they are physically much more realistic
and their structures depart markedly from static models.
Advances are being made both concerning
pulsation models (even including dust formation) which primarily
concentrate on temporal variations of, for instance, mira
variables (see, i.a.,  \cite{bowen}, \cite{fleischer}, \cite{hofner1}, and \cite{hofner2})
and hydrodynamic 3D models which allow for the
formation of spatial inhomogeneities due to convection (c.f.
\cite{frey},
\cite{ludwig},
and others). These are very important frontiers to be explored in order to proceed with
detailed spectroscopy of red giants in general.

Another important aspect when discussing the computation of the
model structure of the atmospheres of cool stars, is the validity of the assumption of
LTE and molecular equilibrium. The assumption of LTE could
lead to erroneous inferences. Hauschildt and collaborators (see, for example, \cite{short})
are working on
codes with which to compute atmospheric structures, assuming statistical equilibrium instead of LTE
to calculate both line and continuum radiative transfer as well as
the gaseous state.

For a large fraction of cool stars, especially evolved AGB stars
and M dwarfs and cooler, dust plays an integral role in
driving winds, where appropriate, and shaping the model structure, for example, by back scattering.
Furthermore, in M dwarfs dust may, for instance, be in a
suspension, heating the line-forming regions \cite{wehrse} or, for even cooler dwarfs, fall to deeper stellar layers.
The study of the process of dust formation and the effect of the dust
on these stars is a field attracting increasing attention (see, for example, \cite{hofner3}, \cite{simis},
\cite{fleischer}, and \cite{hau}).
The examination of K and M dwarfs in this respect is of importance, since they are interesting as
probes of Galactic evolution; they are common and often relatively old, thereby sampling the state of the Milky Way
in earlier phases of its evolution.

Also, magnetic activity on dwarfs should be given some attention.
Magnetic fields could affect the structures of stars, both vertically and horizontally, and certainly affect
the spectra in the infrared through
the Zeeman effect. This sensitivity also opens up another significant aspect of high-resolution, infrared spectroscopy: to measure and map
the magnetic fields on late-type stars, for example, by Doppler imaging (cf. Piskunov's contribution in the
present proceedings). Ultimately, this will also lead to better models of stellar atmospheres and synthetic spectra
for abundance analyses, not least for cool dwarfs which are known to have overall atmospheres and fluxes
considerably affected at activity regions on the stellar surface.

To summarize, all the above-mentioned aspects should, where appropriate, be taken into account
when modelling an atmosphere of a cool star.
For instance, the analysis of atmospheres of M dwarfs needs models superior to the classical ones.
These atmospheres are not well understood due to the presence of a host of molecular lines,
the inhomogeneity and magnetic fields of their atmospheres,
and the presence of dust in their photospheres. Their spectra may also be affected by an
optically thin, H$_2$ convection zone \cite{wehrse}.
Today, only one-dimensional, LTE models, treating convection through the mixing-length
approximation,
exist as a standard for abundance analyses of red giants and dwarfs. Diagnostic tests of such
models are, therefore, very important. Realistic 3-D, non-LTE models are needed
for many investigations, in particular to estimate systematic errors in studies based on standard models.

\subsection{The status of synthetic spectra at $1-5\,\mu$m}

We now turn to the second question on how good we can expect synthetic spectra at $1-5\,\mu$m
to be. Generally it is possible to synthesize spectra at a relatively high accuracy.
The accuracy of the synthetic spectra depends on the input data and the validity of the
assumed approximations of the
physics. Given a model atmosphere, in order to calculate a synthetic spectrum in LTE,
atomic and molecular line-data are required: wavelengths, identifications, excitation energies, transition probabilities
and statistical weights, line-broadening parameters, as well as partition functions and
dissociation energies for the molecules.
For the computation of an atmospheric structure it is necessary to have global absorption data as complete as possible,
but these individual data need not be very accurate.
For a calculation of a synthetic spectrum, on the other hand, the accuracy is especially important for data directly
affecting the spectral diagnostics to be calculated.
For synthetic spectra in the infrared, more and better data
are definitely needed both for atoms and a large number of molecules.
It should also be noted that for pressure broadening of molecular lines, especially in dwarfs, available evidence
indicates that Uns\"old's classical prescription (which is used in current calculations)
is only correct to about one order of magnitude (as for atoms; Barklem, 2003, private communication).
In general, astrophysicists should gratefully acknowledge the important, difficult, and tedious
work performed by atomic and molecular physicists!

If LTE is not assumed, a full, computationally demanding, statistical-equi\-librium
calculation is required, needing (as yet, uncertain or non-existent) collisional and radiative transition
data for all `important' transitions for the modelled atoms, ions, and/or molecules. Simultaneously, the radiation field
throughout the photosphere has to be calculated. For certain cases, a non-LTE description of the line formation may
be crucial. Unfortunately, however, we do not know as yet for which spectral diagnostics this is so.

\section{Some recent examples from the literature}

As of today, there exist some tens of articles in the literature on chemical abundance analyses of cool stars based on
high-resolution spectroscopy
at $1-5\,\mu$m. Here, only a few recent examples will be high-lighted. It is clearly an emerging field, which can be expected to
generate much more scientific information in
the near future, in particular thanks to the realization of high-resolution, infrared spectrometers such as CRIRES.

Several interesting studies have been published based on high-resolution spectra observed
with the Phoenix spectrometer mounted on the Gemini South telescope.
For instance, Cuhna et al. \cite{cuhna} measured the abundance of fluorine based on lines from the HF molecule
for a sample of red giants in the Large Magellanic Cloud and the
Galactic globular cluster $\omega$ Cen.
Based on these observations, performed at $2.3\,\mu$m at $R=50\,000$, they conclude that Asymptotic Giant Branch stars do
not seem to be the chief contributor to
fluorine, after all. Wolf-Rayet stars could be important players in this context.
We note that the Wolf-Rayet stars have been suggested to be a major contributor also for carbon,
again even more significant than the AGB stars \cite{bg}.

A further example of abundance determinations using the Phoenix spectrometer is
the work by Smith et al. \cite{smith}.
They determined abundances of $^{12}$C, $^{13}$C, $^{14}$N, $^{16}$O, Na, Sc, Ti, and Fe in 12 red giants
also in the Large Magellanic Cloud at $R = 50\,000$. The abundance pattern found shows evidence of material characteristic of the
first dredge-up, i.e. mixed material processed by the CN cycle.



Examples of abundance determinations using both Phoenix (then mounted on the Kitt Peak 2.1 m telescope) and the high-resolution NIRSPEC spectrometer
(mounted on a Keck telescope) are the oxygen studies by Mel\'endez et al. \cite{melendez1,melendez2}.
Oxygen abundances for metal-poor Galactic stars were determined from vibration-rotational OH lines at a resolution of
$R = 40\,000 - 50\,000$. A constant, relative oxygen abundance, independent of metallicity [Fe/H], was found to
follow the other $\alpha$-elements:
[O/Fe]$=  0.4$ for  $-2 < $[Fe/H]$ < -1$. This investigation supports the idea that oxygen is
synthesized in Supernovae Type II.

Origila et al. \cite{origlia} determined chemical abundances for four giants in old, metal-rich globular clusters in the Galactic Bulge
using NIRSPEC at a (medium) resolution of $R = 25\,000$. This is an example of utilizing the fact that infrared light can more
easily penetrate dust.
They found a metallicity of [Fe/H]$  =  -0.3\pm0.2$, an oxygen abundance of  [O/H]$  =  0.3 \pm 0.1$, and an $\alpha$-element
enhancement of [$\alpha$/Fe]$ = 0.3 \pm 0.2$ for giants in two different globular clusters. The composition found in the
globular-cluster giants is similar to the one found in field stars in the Bulge.
This seems to support the idea that these clusters were formed early, with a rapid enrichment.

Pilachowski et al. \cite{pila:2} observed four giants in the globular cluster M3
using NIRSPEC at medium resolution ($R=25\,000$) and  Phoenix to determine
C-isotopes. They find typical $^{12}$C/$^{13}$C values for three of the globular cluster giants, but a higher
value for a Li-rich giant, giving support to the idea that a Li-burning shell may be
the cause of the existence of Li-rich giants in globular clusters.

Aoki \& Tsuji \cite{aoki} analyse FTS spectra of K and M giants, observed at KPNO ($R\sim 100\,000$)
in the H, K, and L Johnson bands, to determine nitrogen abundances from
CN and NH lines. They discuss the abundance trends with respect to spectral
type and conclude that extra mixing and additional CN-processing after the 1st dredge-up
is required for the observed giants.

Finally, Viti et al. \cite{viti} presented a metallicity determination of the M dwarf CM Draconis by
measuring CO first overtone lines ($v = 0 - 2$ and $1 - 3$) using CGS4 at UKIRT at a (medium)
resolution of $R = 10\,000$.
The CO bands were modelled and are found, as expected, to vary with metallicity and to
be a powerful diagnostic tool for analyzing M dwarfs.
The CO vibration-rotational lines give a metallicity of [Fe/H]$ = -1$ for CM Dra.

\section{Future prospects for high-resolution, near-infrared spectroscopy of element abundances in cool stars}

There are now, in principle, three different types of possible applications
of the recent developments of $1-5\,\mu$m spectroscopy for determining
stellar element abundances. First, one could study \emph{fainter} objects than before,
for example, reaching for dimmer, nearby dwarfs or stars in external galaxies.
Second, one could study \emph{more} objects, performing systematic studies of
populations or performing surveys of complete samples, e.g. within a certain
volume. Third, one could strive towards \emph{higher accuracy} in the observations
or the analyses, in order to support the element-abundance analyses, with details about
observed atmospheric velocity-fields or magnetic fields.

The limiting factors in these various approaches will be different. In the
first case for dwarfs, the modelling of atmospheres and spectra, plagued by
complex molecular absorption and dust as well as diffusion, surface
activity etc., will most probably limit the accuracy of the abundances much more
than the spectral data as such. For distant giants in the Local Group galaxies,
the analyses will be relatively straight-forward as long as the stars seem
similar to stars known already from the Galaxy or the Magellanic clouds. When
chemical abundances, or other spectral characteristics, seem different or exotic --
in many respects the most interesting case --  the
accuracy in the atmospheric parameters derived will, however,
probably be relatively low, in particular
as long as high S/N and high-resolution spectra cannot be acquired
across wide spectral regions. For the second case, the usefulness of surveys
will be considerable as they will presumably further illuminate the multitude
of different types of AGB stars, and of different degrees of activity on
red dwarfs. They may thus give further clues concerning the role of these
giants in stellar evolution and nucleosynthesis, and regarding the evolution
of stellar magnetic fields and their dynamos.
The great use of surveys of chemical abundances in dwarfs and giants in the spectral
interval F~-~K, e.g. for studies of Galactic chemical evolution, will however
probably  not be correspondingly important for studies in the infrared region of
M and carbon stars, due to the individuality of these latter
stars, essentially reflecting the marked effects caused by molecules and dust, which in turn
depend on differences in element abundances, in particular that of the C, N, and O elements.
Also, the difficulty in determining abundances of elements in dusty, heavily
line-blanketed and inhomogeneous atmospheres, will most probably limit the
efforts. As regards the third application, the more detailed study of the
underlying atmospheric physics is no doubt a very
significant aspect of the new spectroscopic possibilities. First after establishing a
more profound understanding of the physics of cool-star atmospheres, shall we
see further important developments in abundance determinations of elements beyond the mere
empirical inter- or extrapolation from more nearby or bright stars.
For example, in order to
understand the real news, or to be able to distinguish it at all, we need to look
deeper at what determines the phenomena we see, and thought we understood.

The future of chemical-abundance determinations in cool stars seems thus more
interesting than easily predictable. No doubt, however, with CRIRES
and similar spectrometers at large and intermediate-size telescopes,
in combination with more realistic models of cool stars, we may in the end look
forward to reliable abundance determinations in a variety of fundamentally
interesting objects in the future. As a major side effect, or perhaps
as the most rewarding part of the effort, we will learn a lot more about
the physics of cool stars, the interesting interplay between stellar nuclear
reactions, pulsations and travelling shocks, magnetic fields, convection,
mass loss, dust formation and non-equilibrium radiation fields. CRIRES
will be important in bringing our understanding of stars closer to physical
reality.

\bigskip

\noindent \textbf{Acknowledgments} We are grateful to Dr. Jean-Pierre Maillard for valuable help.

%

\end{document}